\documentclass[12pt]{article}
\usepackage[utf8]{inputenc}
\usepackage[english]{babel}
\usepackage[affil-it]{authblk}

\usepackage{amssymb,amsfonts,amsthm,psfrag}
\usepackage{graphicx}
\usepackage{amsmath}

\def\fun#1#2{\lower3.6pt\vbox{\baselineskip0pt\lineskip.9pt
\ialign{$\mathsurround=0pt#1\hfil ##\hfil$\crcr#2\crcr\sim\crcr}}}
\begin{document}

\title{The physics of the $\eta$ - $\eta'$ system versus $B^0 \rightarrow J/\Psi \ \eta (\eta')$ and $B_s \rightarrow J/\Psi \ \eta (\eta')$ decays.}

\author[1,2]{M.A. Andreichikov \thanks{andreichicov@mail.ru}}
\author[3,4]{M.I. Eides \thanks{meides@g.uky.edu}}
\author[1,2,5]{V.A. Novikov \thanks{novikov@itep.ru}}
\author[1,2,5]{M.I. Vysotsky \thanks{vysotsky@itep.ru}}

\affil[1]{A.I. Alikhanov Institute for Theoretical and Experimental Physics, 117218, Moscow, Russia}
\affil[2]{Moscow Institute of Physics and Technology, 141701, Dolgoprudniy, Moscow region, Russia}
\affil[3]{Department of Physics and Astronomy, University of Kentucky, Lexington, KY 40506-0055, USA}
\affil[4]{Petersburg Nuclear Physics Institute, Gatchina, St.Petersburg 188300, Russia}
\affil[5]{National Research University Higher School of Economics, 101978, Moscow, Russia}

\date{}

\maketitle

\begin{abstract}
An approach to the properties of the $\eta$ - $\eta'$ system developed to solve the famous $U(1)$ problem is used to calculate the partial widths ratios  to $\eta$ and $\eta'$ in the  $B^0 \rightarrow J/\Psi \ \eta(\eta',\ \pi^0)$ and  $B_s \rightarrow J/\Psi \ \eta(\eta')$ decays. We obtain the results in agreement with the experimental data.

\end{abstract}

\section{Introduction}
\label{Sec1}

Solution of the $U(1)$ problem is an important achievement of QCD at low energies \cite{1}-\cite{5}. It provides a successful description of the properties of $\eta'$- and $\eta$-mesons. The results obtained in solution of the $U(1)$ problem will be used below to find the ratios of the $B^0 \rightarrow J/\Psi \ \eta$, $B^0 \rightarrow J/\Psi \ \eta'$ and $B^0 \rightarrow J/\Psi \ \pi^0$
decay probabilities as well as the ratio $\Gamma (B_s \rightarrow J/\Psi \ \eta)/ \Gamma (B_s \rightarrow J/\Psi \ \eta') $.

Relative probabilities of the $B^0 \rightarrow J/\Psi \ \eta$ and $B^0 \rightarrow J/\Psi \ \eta'$ decays were measured in \cite{6,7}, while the relative probability of the $B^0 \rightarrow J/\Psi \ \pi^0$ decay was measured in \cite{9a}. The partial widths of the $B_s \rightarrow J/\Psi \ \eta$ and $B_s \rightarrow J/\Psi \ \eta'$ decays were measured in \cite{7,8,9} while the probabilities of the decays with $\psi(2S)$ in the final state were determined  in \cite{7,6p}. In what follows we will use the averaged values of the ratios of these probabilities presented in the Review of Particle Properties \cite{PDG}.

In the case of $B^0$-meson the decays which we are studying occur due to the $\bar{b} \rightarrow c \bar{c} \bar{d}$ quark transition. In the case of $B_s$ the quark transition $\bar{b} \rightarrow c \bar{c} \bar{s}$ is responsible for the decays to the same final states. The $c \bar{c}$ pair  forms $J/\Psi-$ or $\psi(2S)$-meson, while the remaining light quark combines with a spectator quark forming $d \bar{d}$ state in the case of $B^0$ decays or $s \bar{s}$ state in the case of $B_s$ decays.

We will investigate the consequences of the hypothesis that the probability amplitude of the $\eta$-meson production is proportional to the matrix element $\langle 0 | \bar{d} \gamma_5 d | \eta \rangle$ in the case of $B^0$ decay and the matrix element $\langle 0 | \bar{s} \gamma_5 s | \eta \rangle$ in the case of $B_s$ decay. Similar matrix elements with the substitution $\eta\to\eta'$ describe $J/\Psi \ \eta'$ production and with substitution $\eta\to\pi^0$ they describe  $J/\Psi \ \pi^0$ production. In Section \ref{Sec2} we neglect the isotopic symmetry violation. We will discuss possible consequences of the violation of isotopic symmetry in Section \ref{Sec3}.

\section{Estimates of the decay probabilities}
\label{Sec2}

The naive wave functions of the isospin singlet pseudoscalar mesons in the framework of the quark model should be $\pi_1 = \frac{1}{\sqrt{2}}(\bar{u} \gamma_5 u + \bar{d} \gamma_5 d)$ and $\pi_2 = \bar{s} \gamma_5 s$. The mass of $\pi_1$ should not exceed that of $\pi$-meson in a stark contrast with the measured mass of the $\eta$-meson. This is the essence of $U(1)$ problem which in the framework of QCD is resolved due to mixing of the massless in the chiral limit $m_u = m_d = 0$ $\pi_1$-meson with the massless ghost state made from gluons. The state $\pi_2$ mixes with this ghost as well, while the $SU(3)$ octet superposition of $\pi_1$ and $\pi_2$ effectively decouples from the ghost in the limit of light $s$-quark $m_s = m_u = m_d \ll \Lambda_{QCD}$. In this way light $\eta_0 = (\bar{u}\gamma_5 u + \bar{d}\gamma_5 d - 2 \bar{s}\gamma_5 s)/\sqrt{6}$ and heavy $\eta_0' = (\bar{u}\gamma_5 u + \bar{d}\gamma_5 d + \bar{s}\gamma_5 s)/\sqrt{3}$ states are formed which mix due to the heavyness of $s$-quark, $m_s \gg m_u, m_d$. This is the way how physical $\eta$- and $\eta'$-mesons are formed and gluons are very important in this process.

The matrix elements we are looking for were calculated in \cite{5p,5}. They are expressed through the following parameters:

\begin{eqnarray}
  f_1  = f_\pi = 132 \; \mbox{\rm MeV} \; , \;\; f_K = 155 \; \mbox{\rm MeV} \; , \;\; f_2 = 2f_K - f_\pi = 178 \;\mbox{\rm MeV} \; , \label{A} \nonumber \\
 m_1  = m_\pi \; , \;\; m_2^2 = 2 m_K^2-m_{\pi}^2 \; , \;\; \mu_1/\mu_2 = \sqrt{2} f_2/f_1 = 1.91 \; , \\
 m_1^2  + m_2^2 + \mu_1^2 + \mu_2^2 = m_\eta^2 + m_{\eta^\prime}^2 \; , \;\; \mu_1^2 = 0.57 \; \mbox{\rm GeV}^2 \; , \;\; \mu_2^2 = 0.16 \; \mbox{\rm GeV}^2 \; , \nonumber
\end{eqnarray}
 where $\mu_i$ parameterize transition amplitudes of the ghost to $\pi_i$,
 $\langle a_{\nu }|\pi_{1,2}\rangle = -iq_{\nu}\mu_{1,2}$.
According to \cite{5p,5} matrix elements of the divergence of the strange quarks axial current
$P_2 = 2i m_s \bar{s} \gamma_5 s$ are:
\begin{eqnarray}
 \langle 0|P_2|\eta\rangle = -\sqrt{\frac{f_2^2 m_2^4 (m_1^2 + \mu_1^2 - m_\eta^2)}{m_{\eta^\prime}^2 - m_\eta^2}} = -0.056 \; \mbox{\rm GeV}^3 \;\; , \nonumber \\
\langle 0|P_2|\eta^\prime\rangle = \sqrt{\frac{f_2^2 m_2^4 (m_{\eta^\prime}^2 - m_1^2 - \mu_1^2)}{m_{\eta^\prime}^2 - m_\eta^2}} = 0.062 \; \mbox{\rm GeV}^3 \;\; .
\label{2}
\end{eqnarray}

Matrix elements of the isoscalar axial current divergence
$P_1 = i\sqrt 2 [m_u \bar u \gamma_5 u + m_d \bar d \gamma_5 d]$ are \cite{5p,5}:
\begin{eqnarray}
 \langle 0|P_1|\eta\rangle = \sqrt{\frac{f_1^2 m_1^4 (m_2^2 + \mu_2^2 - m_\eta^2)}{m_{\eta^\prime}^2 - m_\eta^2}} = 1.9 \cdot 10^{-3} \; \mbox{\rm GeV}^3 \;\; , \nonumber \\
\langle 0|P_1|\eta^\prime\rangle = \sqrt{\frac{f_1^2 m_1^4 (m_{\eta^\prime}^2 - m_2^2 - \mu_2^2)}{m_{\eta^\prime}^2 - m_\eta^2}} = 1.8 \cdot 10^{-3} \; \mbox{\rm GeV}^3 \;\; .
\label{3}
\end{eqnarray}

Let us assume exact isotopic symmetry and neglect $u$- and $d$-quark mass differences. Then, according to PCAC, the divergence of the isotriplet neutral axial current is proportional to $\pi^0$ field and its matrix element between the $\eta$-meson and vacuum is zero:
\begin{equation}
 \langle0|\bar u \gamma_5 u - \bar d \gamma_5 d|\eta\rangle = 0 \; .
 \label{4}
\end{equation}
Exactly the same relationship holds for $\eta'$. All matrix elements we are looking for in the case of $B^0$ decays can be extracted from (\ref{3}).

We are considering $P$-wave decays and their probabilities are proportional to the third power of the momentum of the produced particles:
\begin{equation}
  |\bar p|^3 \sim \left[1-\left(\frac{\mu+m}{M}\right)^2\right]^{3/2} \left[1-\left(\frac{\mu-m}{M}\right)^2\right]^{3/2} \; .
  \label{5}
\end{equation}
Here $M$ is the mass of the decaying particle ($B_0$ or $B_s$) and $\mu$ and $m$ are the masses of decay products. The numerical values of this factor for the decays under consideration are given in the Table.
 \begin{center}

{\bf Table:} The numerical values of the right-hand part of equation (\ref{5})

\end{center}

$$
\begin{array}{ll}

B^0 \to J/\psi \eta & 0.25 \\
B^0 \to J/\psi \eta^\prime &  0.20 \\
B^0 \to J/\psi \pi^0 &  0.28 \\
B_s \to J/\psi \eta &  0.27 \\
B_s \to J/\psi \eta^\prime &  0.23 \\
B_s \to J/\psi \pi^0 &  0.30 \\
B_s \to \psi(2S)\eta &  0.12 \\
B_s \to \psi(2S)\eta^\prime &  0.08
\end{array}
$$

For the ratios of the decay probabilities in case of $\eta$ and $\eta'$ production we obtain:
\begin{eqnarray}
\frac{{\rm Br}(B^0 \to J/\psi\eta)}{{\rm Br}(B^0 \to J/\psi\eta^\prime)} =
\left(\frac{p_\eta}{p_{\eta'}}\right)^3\left[\frac{\langle 0|P_1|\eta\rangle}{\langle 0|P_1|\eta^\prime\rangle} \right]^2
= 1.39 \ [1.11 \pm 0.47] , \nonumber \\
\frac{{\rm Br}(B_s \to J/\psi\eta)}{{\rm Br}(B_s \to J/\psi\eta^\prime)} =
\left(\frac{p_\eta}{p_{\eta'}}\right)^3\left[\frac{\langle 0|P_2|\eta\rangle}{\langle 0|P_2|\eta^\prime\rangle} \right]^2 = 0.96 \ [1.15 \pm 0.08] ,
\label{6}
\end{eqnarray}
where $p_\eta$ and $p_{\eta'}$ here and in the formulae below are the momenta of the final $\eta$ and $\eta'$ in each of the  respective decays. In the brackets here and in the similar equations below are the results of measurements averaged according to \cite{PDG}.

We use the relationship ($2m_s/(m_u + m_d) = 27.3 \pm 0.7$, see \cite{PDG})
\begin{equation}
\frac{\langle 0|\bar s \gamma_5 s|\eta\rangle}{\langle 0| \bar d \gamma_5 d|\eta\rangle} =
\frac{\langle 0|P_2|\eta\rangle/(2m_s)}{\langle 0|P_1|\eta\rangle/(\sqrt 2 (m_u + m_d))}= -1.53 \pm 0.03
\label{7}
\end{equation}
to determine the ratios of the probabilities of $B_s$ and $B_0$ decays:
\begin{eqnarray}
\frac{{\rm Br}(B_s \to J/\psi\eta)}{{\rm Br}(B^0 \to J/\psi\eta)} =
\left(\frac{p_\eta}{p_{\eta'}}\right)^3\frac{1}{\sin^2\theta_c}\left[\frac{\langle 0|\bar s \gamma_5 s|\eta\rangle}{\langle 0| \bar d \gamma_5 d|\eta\rangle}\right]^2=   \nonumber \\
= 52 \;\; \left[\frac{(4.0 \pm 0.7)\cdot 10^{-4}}{(10.8 \pm 2)\cdot 10^{-6}} = 37\pm 9 \right] \; .
\label{8}
\end{eqnarray}
The factor $\sin^2\theta_c = 0.22^2$ takes into account suppression of $\bar{c}d$ charged current by the sinus of the Cabibbo angle.

The decay $B^0 \rightarrow J/\Psi \ \pi^0$ can be considered similarly. Using the PCAC relationship
\begin{equation}
 i\langle 0| 2m_u \bar u \gamma_5 u - 2m_d \bar d\gamma_5 d |\pi^0 \rangle = \sqrt 2 f_\pi m_\pi^2
 \label{8b}
\end{equation}
and taking into account that $\langle 0| \bar u \gamma_5 u + \bar d \gamma_5 d | \pi^0 \rangle = 0$ we obtain
\begin{equation}
i\langle 0|\bar d \gamma_5 d|\pi^0\rangle = -\frac{f_\pi m_\pi^2}{\sqrt 2(m_u + m_d)} \;\; .
\label{10}
\end{equation}
Then the ratio of the decay probabilities is:
\begin{eqnarray}
\frac{{\rm Br}(B^0 \to J/\psi\pi^0)}{{\rm Br}(B^0 \to J/\psi\eta)} = \left(\frac{p_\eta}{p_{\eta'}}\right)^3 \left(\frac{f_\pi m_\pi^2}{\langle 0|P_1|\eta\rangle}\right)^2 = 1.8 \nonumber \\
\left[\frac{(1.7 \pm 0.1)\cdot 10^{-5}}{(1.08 \pm 0.23)\cdot 10^{-5}} = 1.6 \pm 0.4\right] \; .
\label{11}
\end{eqnarray}
Comparison of the theoretical and experimental (in square brackets) results in eq.(\ref{6}), eq.(\ref{8}), and eq.(\ref{11}) shows a satisfactory agreement.

These decays were analyzed in \cite{7} with the help of the wave functions of $\eta$- and $\eta'$-mesons. Exploiting the observation \cite{B,F} that the gluon admixture in $\eta$ is negligible  the authors of \cite{7} come to the conclusion that gluon admixture in $\eta'$ is small (see also \cite{10}). Let us remind that the large mass of $\eta'$ is explained by the large gluon admixture.
In the $SU(3)$ limit $m_u=m_d=m_s<<\mu_i$ decoupling of $\eta$-meson from gluons really occurs. In this limit instead of eq.\eqref{A} we obtain (we correct some misprints in \cite{5p,5} in the expressions for the $\eta$
and $\eta'$ masses in the $SU(3)$ limit)

\begin{eqnarray}
 f_K=f_{\pi}= f_1 = f_2,\; m_K=m_{\pi}= m_1=m_2,\; \;\mu_1=\sqrt{2}\mu_2, \nonumber \\
 m^2_{\eta}=(m_1^2+2m_2^2)/3=m_{\pi}^2,\; m_{\eta'}^2=\mu_1^2+\mu_2^2=3\mu_2^2
\end{eqnarray}
We see that in the case of exact $SU(3)$ symmetry and tiny quark masses all mass of the $\eta'$-meson is due to coupling with the gluons.

In the real world the $SU(3)$ flavor symmetry is violated and even $\eta$-meson does not decouple from the gluons. The decays  $J/\psi \rightarrow \eta(\eta') \gamma$ were considered in \cite{NSVZ}. The authors calculated the ratio $\langle 0|Q| \eta \rangle /\langle 0 |Q| \eta^\prime \rangle \approx 0.46$ ($Q={\alpha_s}/(8\pi) G \tilde G $). This result is close to $0.36$ obtained in \cite{5p,5}. Using this ratio the ratio of the $J/\Psi$-meson decay probabilities was obtained:

\begin{equation}
\frac{\Gamma(\psi \rightarrow \eta \gamma)}{\Gamma (\psi \rightarrow \eta' \gamma)} = \left| \frac{\langle 0|Q|\eta \rangle}{\langle 0|Q|\eta' \rangle} \right|^2 \left|\frac{{p}_{\eta}}{{p}_{\eta'}} \right|^3 = 0.16 \div 0.25,
\label{12}
\end{equation}
to be compared with the experimental result $[1.10(3)\cdot 10^{-3}]/[5.2(2)\cdot 10^{-3}] = 0.21$ \cite{PDG}. This result confirms large admixture of gluons in the $\eta$-meson.

In the same way as above we calculate the ratio of $B_s\to\psi (2S)\eta(\eta')$ decay probabilities 
\begin{eqnarray}
\frac{{\rm Br}(B_s \to \psi(2S)\eta)}{{\rm Br}(B_s \to \psi(2S)\eta^\prime)} =
\left(\frac{p_\eta}{p_{\eta'}}\right)^3 \left[\frac{\langle 0|P_2|\eta\rangle}{\langle 0|P_2|\eta^\prime\rangle} \right]^2=
1.22   \nonumber \\  \left[\frac{(3.3 \pm 0.9)\cdot 10^{-4}}{(1.29 \pm 0.35)\cdot 10^{-4}} = 2.6 \pm 1\right] \;.
\label{13}
\end{eqnarray}

We also calculate the ratio of the decay probabilities  for the decays $B_s \to \psi(2S)\eta^\prime$ and $B_s \to \psi(2S)\eta$. The probabilities of the charmonium states production in the weak $b \rightarrow c \bar{c}q$ decays are proportional to the $c \bar{c}$
wave function squared at zero. The probabilities of charmonium decay to $e^+ e^-$ pair are also proportional to the wave function squared at zero  and we obtain:
\begin{equation}
\frac{{\rm Br}(B_s \to \psi(2S)\eta^\prime)}{{\rm Br}(B_s \to J/\psi\eta^\prime)} = \left(\frac{p_{\psi(2S)}}{p_{J/\psi}}\right)^3 \frac{\Gamma(\psi(2S) \to e^+ e^-)}{\Gamma(\psi \to e^+ e^-)} = 0.17 \;\; \left[\frac{1.29 \pm 0.35}{3.3 \pm 0.4} = 0.4 \pm 0.1\right] \; ,
\label{133}
\end{equation}
where $p_{\psi(2S)}$ and $p_{J/\psi}$ are the momenta of the final $\psi(2S)$ and $J/\psi$, respectively.

Consider finally radiative decays of the $\phi$-meson with $\eta$ and $\eta'$ in the final state. The ratio of the partial widths  $R_\phi= BR(\phi\rightarrow\eta'\gamma)/BR(\phi\rightarrow\eta\gamma)$ was measured in \cite{13}. Again using the matrix elements from eq.(2) to describe these $P$-wave decays we obtain:

\begin{equation}
R_\phi=\left[\frac{\langle 0|P_2|\eta^\prime\rangle}{\langle 0|P_2|\eta\rangle} \right]^2
\frac{p^3_{\eta'}}{p^3_{\eta}} = 5.4\cdotp10^{-3},
 \label{21}
\end{equation}
to be compared with the experimentally measured ratio $(4.8\pm0.2)\cdotp10^{-3}$ \cite{13}.

\section{The deviations from isotopic symmetry}
\label{Sec3}

We used isotopic symmetry calculating the matrix elements and
now we would like to address corrections due to violation of the isotopic symmetry. There are two sources of isotopic symmetry violation, QED corrections  and $u$- and $d$-quark mass differences. The QED corrections are very small numerically and we will not consider them here. The situation with the quark mass differences is more involved. The corrections of the order of $(m_d - m_u)/m_s$ or $(m_d - m_u)/\Lambda_{QCD}$ are also well below the level of accuracy to which we may pretend. The question is if the corrections of the order of $(m_d - m_u)/(m_d + m_u)$ do exist. They would be important numerically and are interesting from the theoretical point of view.

The difference of the masses of $u$- and $d$-quarks leads to $\eta^0$-$\pi^0$ mixing (in this Section the upper script ``0'' mean the isotopically symmetric case). The $SU(2)$ violating potential in the QCD Hamiltonian  is

\begin{equation} \label{tertpot}
V=\frac{m_u-m_d}{2}(\bar u u-\bar d d).
\end{equation}

We use nonrelativistic perturbation theory to obtain the first order correction to the $\eta$-meson wave function:
\begin{equation}
| \eta \rangle = |\eta^0 \rangle + \frac{\langle\pi_0|V|\eta^0\rangle}{m_{\eta^0}^2 - m_{\pi^0}^2} |\pi^0 \rangle  \approx |\eta^0 \rangle + \frac{\sqrt{3}}{4} \frac{m_d - m_u}{m_s} | \pi^0 \rangle,
\label{15}
\end{equation}
where we used the soft-pion theorem to calculate ($\langle\bar uu\rangle=\langle\bar d d\rangle=\langle\bar s s\rangle\approx (-250~MeV)^3$ is the $SU(3)$ symmetric quark condensate)

\begin{equation}
\langle\pi_0|V|\eta^0\rangle=-\frac{1}{f_{\pi}^2} \frac{m_d - m_u}{\sqrt{3}}\langle\bar uu+\bar d d\rangle,
\qquad m_{\eta^0}^2 - m_{\pi^0}^2\approx -\frac{1}{f^2_{\pi}}\frac{8m_s}{3}\langle\bar ss\rangle,
\end{equation}
for more details see \cite{5p,5,11p}

Then the correction to the matrix element under discussion is 
\begin{eqnarray}
  \langle 0| \bar{d} \gamma_5 d | \eta \rangle =  \langle 0| \bar{d} \gamma_5 d | \eta^0 \rangle + \frac{\sqrt{3}}{4} \frac{m_d - m_u}{m_s}  \langle 0 | \bar{d} \gamma_5 d | \pi^0 \rangle = \nonumber \\
 =  \langle 0 | \bar{d} \gamma_5 d | \eta^0 \rangle \left[1 + O\left(\frac{m_d - m_u}{m_s}\right) \right].
\label{16}
\end{eqnarray}
First order correction to the $\pi$-meson wave function is
\begin{equation}
|\pi \rangle = |\pi^0 \rangle - \frac{\langle\pi_0|V|\eta^0\rangle}{m_{\eta^0}^2 - m_{\pi^0}^2}|\eta^0 \rangle  = |\pi^0 \rangle - \frac{\sqrt{3}}{4}\frac{m_d - m_u}{m_s}|\eta^0 \rangle,
\label{17}
\end{equation}
and similarly to (\ref{16}) we obtain a very small correction to the matrix element
\begin{equation}
  \label{18}
   \langle 0 | \bar{d} \gamma_5 d |\pi \rangle =  \langle 0| \bar{d} \gamma_5 d|\pi^0 \rangle \left[ 1 + O \left(\frac{m_d - m_u}{m_s} \right) \right].
\end{equation}
Now we can also estimate the relative probability of the $B_s \rightarrow J/\Psi \ \pi$ decay
\begin{equation}
  \label{20}
  \frac{{\rm Br} (B_s \rightarrow J/\Psi \pi)}{{\rm Br} (B_s \rightarrow J/\Psi \eta)} = \left(\frac{p_\pi}{p_{\eta}}\right)^3 \frac{3}{16} \left(\frac{m_d - m_u}{m_s} \right)^2 \approx 1.5 \cdot 10^{-4},
\end{equation}
where we used (\ref{17}) to calculate
\begin{equation}
  \label{19}
   \langle 0 | \bar{s} \gamma_5 s | \pi \rangle = -\frac{\sqrt{3}}{4} \frac{m_d - m_u}{m_s}  \langle 0| \bar{s} \gamma_5 s | \eta \rangle.
\end{equation}

\section{Conclusions}
\label{Sec4}

We considered above the $B^0 (B_s) \rightarrow J/\Psi \ (\eta,\eta',\pi^0)$ decays and described the qualitative pattern of these decays using the methods developed in the late 1970s and in early 1980s for solution of the $U(1)$ problem. Moreover, these methods allowed us to obtain quantitative description of the ratios of the partial widths that is in agreement with the experimental data.

We are grateful to A.E.Bondar for bringing the KLOE measurement \cite{13} to our attention.

M.A., V.N. and M.V. are supported by RSF grant 19-12-00123, M.E. is supported by the NSF grant PHY-1724638.

.

\end{document}